\documentclass[referee]{aa}

\usepackage{graphics,natbib}
\input{psfig.sty}

\begin{document}

\title{$H$-band observations of the Chandra Deep Field South%
\footnote{Based on observations collected at the European Southern
Observatory, Chile under programs 66.A-0451 and 68.A-0375}}

\author{Emmanuel Moy\inst{1}, Pauline Barmby\inst{2}, Dimitra Rigopoulou\inst{1},
Jia-Sheng Huang\inst{2}, 
Steven P. Willner\inst{2}, Giovanni G. Fazio\inst{2}}
\institute{
\inst{1} Max-Planck-Institut f\"{u}r extraterrestrische Physik (MPE),
Postfach 1312, Garching D-85741, Germany \\
\inst{2} Harvard-Smithsonian Center for Astrophysics, 60 Garden St.,
MS 65, Cambridge MA 02138}

\authorrunning{Moy et al.}

\titlerunning{$H$-band observations of the CDF-S}

\date{}

\abstract{ We report preliminary results of our $H$-band survey of the Chandra
Deep Field South (CDFS).  The observations were made using SofI on the
NTT, and cover 0.027 square degrees with a  50\,\% completeness limit
of $H$\,=20.5,  and 0.17 square degrees with a 50\,\% completeness
limit of $H$\,=19.8.  We used SExtractor to extract sources from our
fields. In total we have detected 4819 objects. Star-galaxy
separation was performed using the SExtractor parameter ``stellarity
index''. All objects with an index of 0.5 or lower were classified as 
galaxies. According to this criterion, 80 \% of our detections are
galaxies. We then compare our results with previous
observations of the CDFS. Our astrometric solutions are in good
agreement with the Las Campanas Infrared Survey (LCIRS),  the COMBO-17
and the ESO-EIS surveys. The photometry of our catalog compares
satisfactorily with the results of the LCIRS, as well as with the
GOODS data. Galaxy number counts are presented and compared with the
LCIRS results. The present data are intended to complement the recent and
future multi-wavelenghth observations of the CDFS and will be used, 
in conjuction with additional multiband photometry, to find counterparts
of the upcoming mid-infrared surveys with SIRTF.

}

\maketitle

\section{Introduction}
One of the many reasons for performing a deep survey is to study
large-scale structure at high redshifts ($z\geq 1$) and to directly
probe  galaxy evolution {\em from early epochs to present days}.  Such studies
allow us to trace the history of galaxy formation and evolution
\citep{ellis, cohen} and to improve our understanding of the sources
of the various components of the cosmic background
\citep{totani,madau}. High redshift galaxies are often quite faint,
and the use of spectroscopy to measure redshifts and other properties
can be prohibitively expensive. The alternative way to estimate
redshifts for large samples is to apply photometric techniques
\citep{zphot1, zphot2, zphot3, zphot4}. This approach relies on  the
detection of typical SED features. Optical (UBVRI) data are able to
detect the Balmer Break at 4000 \AA~  up to $z$\,$\sim$\,1.3, and  the
Lyman Break at 900 \AA~ for $z$\,$\ge$\,3 \citep{lymb}.   For
1.3\,$\le$\,$z$\,$\le$\,3, the  Balmer Break is redshifted  beyond the
$I$ band  and near infrared (NIR) data in the J and H bands are
necessary to constraint the redshift with a high enough accuracy.

Currently, one of the best studied fields in the sky is the Chandra
Deep Field South (CDF-S),  located at J2000 coordinates $03^{\rm
h}32^{\rm m}28^{\rm s}$, $-27^{\circ} 48\arcmin30\arcsec$
\citep{cdfs1}. The CDF-S is a ``blank'' region of the southern sky
chosen as the target for an extremely deep exposure with the {\em
Chandra} X-ray Observatory.  It has been the subject of observations
with other space-based telescopes as well, including the Hubble Space
Telescope \citep{cdfs_hst1}, as well as with numerous ground-based
telescopes in the optical \citep{eis_opt, combo}, near-infrared
\citep{sar01, lcirs3}, and radio \citep{kel00}.  Several sets of mid-
and far-infrared observations with the upcoming Space Infrared
Telescope Facility \citep[SIRTF;][]{sirtf} are planned, including
Guaranteed Time Observations by the IRAC and MIPS instrument teams and
Legacy program observations by the GOODS \citep{goods} and SWIRE
\citep{swire} projects.

The ESO Imaging Survey (EIS) has performed the most extensive
optical-NIR coverage of the CDF-S to date. It includes $UBVRI$
observations  covering 0.25 square degrees \citep{eis_opt} and $JK_s$
observations  covering $\sim$400 square arcminutes
\citep{eis_ir}. Filling the gap between these two  filters is required
by photometric redshift determination techniques, and implies
observations of the CDFS in the H band. The COMBO-17 survey
\citep{combo} has imaged a $31.5\arcmin \times$30\arcmin\ field
covering the CDF-S in 5 broad-band and 12 narrow-band filters, but
with  a wavelength coverage limited to the interval between the 3650
and 9140 \AA. The Las Campanas Infrared Survey \citep{lcirs2, lcirs3}
has recently performed $H$-band observations covering
$25\arcmin\times$25\arcmin\ in the CDF-S, down to a 5 $\sigma$
limiting magnitude of $\sim$\,20.5. ; however, their field is not
centered on the {\em Chandra} observations.  We note that the GOODS
SIRTF Legacy team is observing a smaller ($10\arcmin\times 15\arcmin$)
area of the CDF-S to a much greater depth than our observations, using
ISAAC on the VLT. Their first data are  available on the ESO web site.

We have conducted an $H$-band survey of the CDF-S, centered on the
X-ray field, and encompassing the  spatial coverage of the EIS
observations as well as the future IRAC pointings. {\em In the present
paper we present the results of this survey.}  The observations and
data reduction are described in Section~\ref{obs};
Section~\ref{source}  presents the source extraction method and the
resulting source lists. The survey performance is  discussed in
Section~\ref{perf}. Our conclusions are summarized in
Section~\ref{conc}.

\section{Observations and Data Reduction\label{obs}}

\subsection{Observations}

We carried out the observations using the SofI near infrared imaging
camera \citep{mcl98} installed on the ESO--NTT telescope at La Silla,
during two observing runs: 15--18 November 2000 and 27--30 November
2001. The $H$-band filter of the SofI imaging mode is centered at
1.653~$\mu$m and is 0.297~$\mu$m wide. The SofI detector is a Hawaii
HgCdTe 1024x1024 array. Sky conditions were photometric throughout the
acquisition of the images with the seeing varying between
0.4--0.8\arcsec.  We defined 32 $4.9\arcmin \times 4.9 \arcmin$ fields
covering the CDF-S area  (see Figure~\ref{fig-pretty})  and observed
each of them in the ``large field'' configuration, with a scale of
0.288\arcsec~pixel$^{-1}$.

The observations were carried out using the ``jitter'' technique, one
of the most efficient methods of sky acquisition with a minimum loss
of observing time. The telescope is offset  by small amounts around a
central position. Offsets are generated randomly but are restricted
within a square box. In our case the box width was 35\arcsec.  A
typical acquisition in jitter mode consisted of 60 1 minute long
frames, reaching a total integration time of 1 hr.  For the four
central fields we repeated the sequence an additional three times, for
a total of 4 hours of integration time. Infrared standard stars from
the \citet{per98} list were observed every hour at similar airmasses
to the science exposures.

\subsection{Sky subtraction}

Data reduction included the usual steps of flat-fielding (based on
dome-flats) and bad pixel fixing, followed by a two-pass
sky-subtraction.  This method prevents bright objects from affecting
the residual sky levels and thus the photometry of faint objects.  The
{\sc jitter} routine in {\sc eclipse} \citep{jitter, eclipse} was used
to make the first-pass sky-subtraction and image registration.  {\sc
jitter} does not have an object-masking capability, so we used the
{\sc dimsum} package in IRAF%
\footnote{IRAF is distributed by the National Optical Astronomy
Observatories, which are operated by the Association of Universities
for Research in Astronomy, Inc., under cooperative agreement with the
National Science Foundation.} to co-add the first-pass images, derive
an object mask from the co-added image, and redo the sky subtraction
with objects masked.  We then combined the final sky-subtracted images
using {\sc jitter} (which is much faster than IRAF-based tools).  

\subsection{Astrometry}

We performed astrometric calibrations of the images using stars taken
from the United States Naval Observatory \citep{usno} and 2 Micron All
Sky Survey (2MASS) databases.  The astrometric solutions were computed
using the tasks {\tt ccmap} and {\tt ccsetwcs} in the IRAF  {\tt
imcoords} package.

In order to check the accuracy of the astrometry our catalogs were
compared with the source lists  extracted from previous observations
of the CDF-S. The relative offsets between the most likely
counterparts of our $H$-band detections in the other source lists were
computed, and  the result is  shown in Fig.~\ref{astrometry}.  There
is a negligible offset in right ascension and an offset of
0.15\arcsec\ in declination between our data and the Las Campanas
Infrared Survey (LCIRS) catalog. The rms  in the difference is $\le
0.24$\arcsec\ in both directions. The comparison is particularly good
with the result of the COMBO-17 survey. The offsets are quite small
($\le 0.02$\arcsec) in both  directions, and the rms is only $\sim
0.2$\arcsec. \cite{eis_opt} found an offset of $\sim 0.2$\arcsec\  in
$\alpha$ and $\delta$ between the EIS and COMBO images. Not
surprisingly, since COMBO  astrometric calibrations are very similar
to ours, we find a similar shift between our data and the  EIS
catalogs, $\sim 0.35$\arcsec, with a rms of 0.2\arcsec.

\subsection{Photometric Calibration}

We measured the instrumental magnitude of the standard stars in a
10-pixel (2.9\arcsec)  radius aperture. The magnitude zeropoint of the
corresponding exposure was computed  as the  difference between the
instrumental and the cataloged $H$-band magnitudes. The zeropoints are
compatible with other results listed on the NTT/SofI web page. Typical
changes in the zeropoint over a night are $\le 0.04$~mag; since this
includes real changes with the airmass, this value is probably an
upper bound on the zeropoint error.  For each CDF-S field (except the
four central ones, see below),  we took the average of the two values
derived from the bracketing  standard star observations as the
zeropoint.  For the co-added images of the four central fields, the
median of the zeropoints computed for  the individual exposures was
adopted.

We tested our photometric calibrations by comparing the magnitudes
measured in a similar aperture  by the LCIRS team and in the present
work. The result is shown in Fig.~\ref{photo}. Our magnitudes  are
0.13~mag brighter than those in LCIRS, with an rms of 0.27 for $14 \le
H \le 19$. The cause  of this difference is not known. We also checked
our measurements by doing photometry on CDF-S H-band images made
available by the GOODS team, although the zeropoint for these data is
uncertain. We find our aperture magnitudes to be 0.06~mag fainter than
those measured on the GOODS images, again with an rms of 0.27
(Fig. \ref{goods}).

\section{Source list\label{source}}

Object detection was performed using version 2.2.2 of the SExtractor
package \citep{ba96}. We performed one SExtractor per field. Each
individual catalog was cleaned from spurious detections in the
outermost parts of the  image, which do not have a full 60-minute
exposure time. Thanks to a partial overlap (30 \arcsec) between our
pointings, these regions are covered to full depth in adjacent fields,
except on the outside of the survey region.

For the extraction we used a 7$\times$7 Gaussian filter with a FWHM of
3 pixels. Weight maps were computed from the image background to
improve the quality  of the source extraction by taking into account
the image quality at each pixel.  The SExtractor detection threshold
was fixed to a conservative value of 3, and the minimum number of
connected pixels above detection threshold to 5 for all fields. The
initial catalogs were then cleaned  of objects near saturated$/$hot
pixels and objects where a reliable estimate of the flux was not
feasible.

The final catalog contains 4819 objects, including 956 detections in
the four deeper central fields.  As an example, the first 40  entries
of the final catalog are presented in Table \ref{tab-hband}.
\footnote{The full catalog is available in electronic form only.}  The
following quantities are listed:
\begin{itemize}

\item Column 1: identification number.

\item Column 2 and 3: right ascension and declination (J2000).

\item Column 4-7: automatic and isophotal magnitudes and respective
errors, as estimated by SExtractor.

\item Column 8-9: aperture magnitude (10 pixels, or 2.9\arcsec\
diameter) and error.

\item Column 10: SExtractor stellarity index.

\end{itemize}

\section{Survey performance\label{perf}}

\subsection{Star-galaxy separation}

Star/galaxy separation in deep observations is always challenging.
{\em Commonly used techniques are based on object morphology and/or
colour properties (e.g. Gardner et al. 19XX, Huang et al., 1997)
when available.}
The SExtractor ``stellarity index'' (CLASS\_STAR) is computed for a
given object from the comparison  of its luminosity profile with the
natural ``fuzziness'' of the image, as estimated from the  seeing
FWHM. CLASS\_STAR is equal to 0 for confirmed galaxies, and to 1 for
confirmed stars. This quantity   is known to be a relatively robust
galaxy identifier  and has the advantage of being widely-used. 
{Our star/galaxy separation is based on the stellarity index.
In Figure~\ref{class_star} we
plot the distribution of the CLASS\_STAR parameter for three different 
magnitude bins.  For the brightest magnitude bin, the
objects are concentrated at the two ends of the histogram, while in
the faintest bin, the distribution is more uniform. 
Thus, we consider hereafter every object with CLASS\_STAR\,$\le
0.5$ or fainter than  $H=19.75$ (outer fields) or $H=20.25$ (inner
fields) as a galaxy.

{\em We note that our adopted value for the star/galaxy separation is
a  very conservative one. For instance Maiahara et al. (2000) have
used  a CLASS\_STAR value of 0.8 (including small magnitude
adjustments at the faint end)  while the EIS classification was based
on a CLASS\_STAR parameter value of 0.9 (Arnouts et al. 2001). As is
obvious from Figure~\ref{class_star} our classification ensures that
the galaxy sample does not suffer from possible contamination by
misidentified stars.}

\subsection{Limiting magnitude}

To estimate the completeness of our catalog, we used the standard
procedure of trying to detect  artificial objects inserted in the
images. The IRAC `mkobject' task was used {\em to simulate and add }
40 stars and 40 galaxies to each field.  The object magnitudes were
distributed according to a shallow power law from 16.5 to 22.5
(17--23.5 for the central fields).  The galaxy half-light radii were
either 0.3\arcsec\ (25\% of the galaxies, chosen at random) or
0.7\arcsec; we found that this mix reproduced the distribution of
(SExtractor parameters) FWHM vs. MAG\_BEST found for our detected
objects.  The size of the PSF used to generate the stars and convolve
with the galaxies was matched to the measured seeing in each field.
Finally, we ran SExtractor on the resulting image, using exactly the
same parameters as for real objects. The entire procedure was repeated
50 times for each field.

Fig.~\ref{completeness} plots the histogram of our completeness
estimation (ie. number of  artificial objects added/ number of
artificial objects detected) for a few fields. We estimated the 50\%
completeness level for each  image by spline-fitting the completeness
histograms; for most of the outer fields, the 50\% completeness level
was $H=19.8$. A few fields --- generally those observed at higher
airmass or in poorer seeing --- had 50\% completeness  levels up to
0.5 mag brighter. The central fields were complete to fainter
magnitudes: $H=20.4$ for fields 19 and 20, and $H=20.7$ for fields 13
and 14.

\subsection{Number Counts}

We generated a completeness curve for each of the two regions as a sum
of the completeness for the individual fields, weighted by the number
of galaxies detected in each field.  The galaxy counts from the
combined catalogs are corrected for incompleteness using this curve,
although in practice this is important only for the last 1 or 2
bins. Figure~\ref{numc} shows the counts for the outer  and inner
fields, together with comparison data from other groups.  Our galaxy
counts match those of \citet{lcirs3} extremely well; the counts of
\citet{mar01} are somewhat higher, and \citet{lcirs3} attribute this
to an incorrect magnitude correction on the part of \citet{mar01}.

\section{Summary\label{conc}}
We have performed an $H$-band survey of the Chandra Deep Field South
covering 0.17  square degrees down to a limiting magnitude of $H_{\rm
Vega}\sim20$. The resulting catalog includes more than 3800
galaxies. We find that our astrometry and photometry  are in generally
good agreement with other surveys in the region, and our derived
galaxy number counts agree well with results from other surveys. \\
The present work fills the gap between the {\em $J$ and $K$ band data
in the EIS database} of the CDF-S, allowing the determination of
accurate photometric redshifts for a large range of $z$.  When
combined with already existing optical data, the present data will
primarily be used to  filter out low-z sources in preparation for the
upcoming mid-infrared  SIRTF surveys on the CDFS. \\

\begin{acknowledgements}

EM acknowledges support of the EU TMR Network ``Probing the Origin of
the Extragalactic  Background Radiation''.  This publication makes use
of data products from the Two Micron All Sky Survey,  which is a joint
project of the University of Massachusetts and the Infrared Processing
and Analysis Center/California Institute of Technology, funded by the
National Aeronautics and Space Administration and the National Science
Foundation.

\end{acknowledgements}
\bibliographystyle{aas}

\begin{thebibliography}{10}
\expandafter\ifx\csname natexlab\endcsname\relax\def\natexlab#1{#1}\fi

\bibitem[{{Arnouts} {et~al.}(2001)}]{eis_opt}
{Arnouts}, S., Vandame, B., Benoist, C.,
 Groenewegen, M. A. T., da Costa, L.,
 Schirmer, M., Mignani, R. P., Slijkhuis, R.,
 Hatziminaoglou, E., Hook, R., Madejsky, R.,
 Rité, C., Wicenec, A. 2001, \aap, 379, 740

\bibitem[{Bertin \& Arnouts(1996)}]{ba96}
Bertin, E. \& Arnouts, S. 1996, \aaps, 117, 393

\bibitem[{{Cohen} {et~al.}(2000)}]{cohen}
Cohen, J.G., Hogg, D., Blandford, R., Cowie, L.L., Hu, E., Songaila, A., 
Shobell, P. \& Richberg, K. 2000, \apj, 539, 29

\bibitem[{{Chen} {et~al.}(2002)}]{lcirs3}
{Chen}, H.-W. {et~al.} 2002, \apj, 570, 54

\bibitem[{Devillard(1997)}]{eclipse}
Devillard, N. 1997, The Messenger, 87, 19

\bibitem[{Devillard(1999)}]{jitter}
Devillard, N. 1999, in ``Astronomical Data Analysis Software and Systems
VII'' eds. Mehringer, D.M., Plante, R.L. \& Roberts, D.A., p.333

\bibitem[{{Dickinson} {et~al.}(2002)}]{goods}
Dickinson, M., Giavalisco, M., et~al., 2002, to appear in the  Proceedings of the ESO Workshop 
``The Mass of Galaxies at Low and High Redshift'', eds. R. Bender and A. Renzini (astro-ph/0204213)

\bibitem[{Ellis(2001)}]{ellis}
Ellis, R.S. 2001, \pasp, 113, 515

\bibitem[{{Fanson} {et~al.}(1998)}]{sirtf}
Fanson, J., Fazio, G., Houck, J., Kelly, T., Rieke, G., Tenerelli, D., Whitten, M. 
1998, in Bely P. Y., Breckinridge J. B., eds, Proc. SPIE 3356, Space Telescopes and Instruments. 
SPIE, Bellingham, WA, p. 478

\bibitem[{{Fern\'andez-Soto}  {et~al.}(1999)}]{zphot3}
Fern\'andez-Soto, A., Lanzetta, K.~M., Yahill, A. 1999, \apj, 513, 34

\bibitem[{{Firth} {et~al.}(2002)}]{lcirs2}
{Firth}, A.~E. {et~al.} 2002, \mnras, 332, 617

\bibitem[{{Fontana} {et~al.}(2000)}]{zphot2}
{Fontana}, A., D'Odorico, S., Poli, F., Giallongo, E., Arnouts, S., Cristiani, S., Moorwood, A., 
Saracco, P. 2000, \aj, 120, 2206

\bibitem[{{Franceschini} {et~al.}(2002)}]{swire}
{Franceschini}, A., Lonsdale C. {et~al.} 2002, to appear in the Proceedings of the ESO Workshop 
``The Mass of Galaxies at Low and High Redshift'', eds. R. Bender and A. Renzini (astro-ph/0202463)

\bibitem[{{Giacconi} {et~al.}(2001){Giacconi}, {Rosati}, {Tozzi}, {Nonino},
  {Hasinger}, {Norman}, {Bergeron}, {Borgani}, {Gilli}, {Gilmozzi}, \&
  {Zheng}}]{cdfs1}
{Giacconi}, R., {Rosati}, P., {Tozzi}, P., {Nonino}, M., {Hasinger}, G.,
  {Norman}, C., {Bergeron}, J., {Borgani}, S., {Gilli}, R., {Gilmozzi}, R., \&
  {Zheng}, W. 2001, \apj, 551, 624

\bibitem[{{Kellermann} {et~al.}(2000){Kellermann}, {Fomalont}, {Rosati},
  {Shaver}, \& {CDF-S Collaboration}}]{kel00}
{Kellermann}, K.~I., {Fomalont}, E.~B., {Rosati}, P., {Shaver}, P., \& {CDF-S
  Collaboration}. 2000, in American Astronomical Society Meeting, Vol. 197,
  9002

\bibitem[{Koo(1985)}]{zphot1}
{Koo}, D.~C. 1985, \aj, 90, 418

\bibitem[{{Le Borgne} {\& Rocca-Volmerange}(2002)}]{zphot4}
Le Borgne, D., Rocca-Volmerange, B., 2002, \aap, 386, 446

\bibitem[{{Madau} {\& Pozzetti}(2000)}]{madau}
Madau, P. \& Pozzetti, L. 2000, \mnras, 312, 9

\bibitem[{Martini(2001)}]{mar01}
Martini, P. 2001, \aj, 121, 598

\bibitem[{Monet(1998)}]{usno}
Monet, D. 1998, \baas,  30, 1427

\bibitem[{Moorwood, Cuby \& Lidman(1998)}]{mcl98}
Moorwood, A., Cuby, J.-G. \& Lidman, C. 1998, The Messenger, 91, 9 

\bibitem[{Persson {et~al.}(1998)Persson, Murphy, Krzeminski, Roth, \&
  Rieke}]{per98}
Persson, S.~E., Murphy, D.~C., Krzeminski, W., Roth, M., \& Rieke, M.~J. 1998,
  \aj, 116, 2475

\bibitem[{{Saracco} {et~al.}(2001){Saracco}, {Giallongo}, {Cristiani},
  {D'Odorico}, {Fontana}, {Iovino}, {Poli}, \& {Vanzella}}]{sar01}
{Saracco}, P., {Giallongo}, E., {Cristiani}, S., {D'Odorico}, S., {Fontana},
  A., {Iovino}, A., {Poli}, F., \& {Vanzella}, E. 2001, \aap, 375, 1

\bibitem[{{Schreier} {et~al.}(2001)}]{cdfs_hst1}
{Schreier}, E.~J. {et~al.} 2001, \apj, 560, 127

\bibitem[{{Steidel} {et~al.}(1995)}]{lymb}
Steidel, C. C., Pettini, M., Hamilton, D. 1995, \aj, 110, 2519

\bibitem[{{Totani} {et~al.}(2001)}]{totani}
Totani, T., Yoshii, Y., Iwamuro, F., Maihara, T., Motohara, K. 
2001, \apj, 550, L137

\bibitem[{{Vandame} {et~al.}(2001)}]{eis_ir}
 Vandame,  B., Olsen,  L.F., J{\o}rgensen, H.E., Groenewegen, M. A. T., Schirmer,  M., Arnouts, S., Benoist, C., da Costa, L., Mignani, R. P., Rit\'{e,} C., Slijkhuis, R., Hatziminaoglou, E., Hook, R., Madejsky, R., Wicenec, A., 2001 (astro-ph/0102300, under revision)

\bibitem[{{Wolf} {et~al.}(2001){Wolf}, {Dye}, {Kleinheinrich}, {Meisenheimer},
  {Rix}, \& {Wisotzki}}]{combo}
{Wolf}, C., {Dye}, S., {Kleinheinrich}, M., {Meisenheimer}, K., {Rix}, H.-W.,
  \& {Wisotzki}, L. 2001, \aap, 377, 442

\end{thebibliography}

\clearpage

\begin{table*}
\caption{First 40 entries of the $H$-band source list. All magnitudes are in the Vega system \label{tab-hband}}
\small
\vskip 0.5 cm
\begin{tabular}{cccccccccc}
\hline

\# & $\alpha$ & $\delta$ & $m_{auto}$ & $\delta m_{auto}$ & $m_{iso}$ & $\delta m_{iso}$ &  $m_{aper}$ & $\delta m_{aper}$ & Class  \\
 & & & & & & & & & \\
    1 & 03:31:32.67 & 27:39:37.19 & 19.44 & 0.082 & 20.33 & 0.059 & 19.47 & 0.069 & 0.01 \\
   2 & 03:31:35.81 & 27:39:29.39 & 19.72 & 0.083 & 20.76 & 0.069 & 19.75 & 0.077 & 0.00 \\
   3 & 03:31:32.74 & 27:35:19.07 & 19.31 & 0.053 & 19.65 & 0.038 & 19.29 & 0.045 & 0.00 \\
   4 & 03:31:35.02 & 27:35:04.61 & 19.37 & 0.051 & 19.54 & 0.034 & 19.38 & 0.053 & 0.16 \\
   5 & 03:31:41.52 & 27:35:05.13 & 19.78 & 0.068 & 20.02 & 0.043 & 19.73 & 0.076 & 0.60 \\
   6 & 03:31:29.99 & 27:35:22.20 & 19.58 & 0.068 & 20.02 & 0.042 & 19.59 & 0.058 & 0.00 \\
   7 & 03:31:31.95 & 27:35:22.88 & 17.86 & 0.019 & 17.99 & 0.013 & 18.02 & 0.014 & 0.02 \\
   8 & 03:31:28.27 & 27:35:24.90 & 17.81 & 0.013 & 17.86 & 0.010 & 17.84 & 0.012 & 0.47 \\
   9 & 03:31:27.16 & 27:35:27.22 & 13.19 & 0.000 & 13.17 & 0.000 & 13.25 & 0.000 & 0.98 \\
  10 & 03:31:40.40 & 27:35:24.21 & 19.99 & 0.084 & 20.58 & 0.055 & 19.94 & 0.085 & 0.27 \\
  11 & 03:31:29.58 & 27:35:29.29 & 16.35 & 0.004 & 16.37 & 0.003 & 16.39 & 0.003 & 0.98 \\
  12 & 03:31:25.50 & 27:35:29.82 & 19.64 & 0.091 & 20.89 & 0.082 & 19.65 & 0.104 & 0.72 \\
  13 & 03:31:36.00 & 27:35:34.12 & 15.15 & 0.001 & 15.14 & 0.001 & 15.17 & 0.001 & 1.00 \\
  14 & 03:31:32.33 & 27:35:35.10 & 19.61 & 0.064 & 19.95 & 0.038 & 19.64 & 0.061 & 0.02 \\
  15 & 03:31:37.93 & 27:35:41.06 & 18.25 & 0.023 & 18.39 & 0.017 & 18.35 & 0.019 & 0.02 \\
  16 & 03:31:40.07 & 27:35:44.24 & 18.09 & 0.026 & 18.39 & 0.019 & 18.35 & 0.020 & 0.01 \\
  17 & 03:31:29.16 & 27:35:49.17 & 19.61 & 0.078 & 20.89 & 0.072 & 19.67 & 0.062 & 0.00 \\
  18 & 03:31:40.11 & 27:35:51.96 & 15.97 & 0.003 & 15.97 & 0.003 & 15.99 & 0.002 & 0.98 \\
  19 & 03:31:36.93 & 27:35:53.85 & 17.99 & 0.015 & 18.05 & 0.012 & 18.00 & 0.014 & 0.99 \\
  20 & 03:31:41.53 & 27:35:55.89 & 19.75 & 0.074 & 20.54 & 0.057 & 19.73 & 0.070 & 0.00 \\
  21 & 03:31:27.31 & 27:35:57.41 & 19.93 & 0.085 & 21.52 & 0.095 & 19.98 & 0.084 & 0.00 \\
  22 & 03:31:29.34 & 27:36:01.93 & 19.20 & 0.062 & 20.34 & 0.053 & 19.49 & 0.052 & 0.00 \\
  23 & 03:31:26.57 & 27:36:03.81 & 19.32 & 0.068 & 20.07 & 0.047 & 19.48 & 0.061 & 0.06 \\
  24 & 03:31:27.36 & 27:36:06.38 & 19.25 & 0.067 & 20.42 & 0.057 & 19.51 & 0.055 & 0.00 \\
  25 & 03:31:35.87 & 27:36:10.69 & 14.96 & 0.001 & 14.96 & 0.001 & 14.99 & 0.001 & 1.00 \\
  26 & 03:31:31.18 & 27:36:12.70 & 18.88 & 0.048 & 19.40 & 0.033 & 19.10 & 0.038 & 0.00 \\
  27 & 03:31:25.98 & 27:36:18.18 & 19.55 & 0.082 & 20.54 & 0.066 & 19.61 & 0.079 & 0.00 \\
  28 & 03:31:39.31 & 27:36:17.86 & 20.51 & 0.101 & 21.31 & 0.074 & 20.41 & 0.128 & 0.60 \\
  29 & 03:31:28.53 & 27:36:19.57 & 18.63 & 0.028 & 18.76 & 0.020 & 18.66 & 0.024 & 0.02 \\
  30 & 03:31:32.00 & 27:36:22.13 & 20.45 & 0.070 & 20.86 & 0.056 & 20.53 & 0.139 & 0.69 \\
  31 & 03:31:32.23 & 27:36:23.73 & 19.33 & 0.052 & 19.66 & 0.032 & 19.36 & 0.048 & 0.01 \\
  32 & 03:31:29.29 & 27:36:24.05 & 18.72 & 0.038 & 19.05 & 0.025 & 18.85 & 0.029 & 0.00 \\
  33 & 03:31:27.76 & 27:36:25.23 & 20.29 & 0.088 & 21.14 & 0.067 & 20.25 & 0.106 & 0.21 \\
  34 & 03:31:36.60 & 27:36:26.34 & 17.67 & 0.016 & 17.83 & 0.012 & 17.81 & 0.012 & 0.03 \\
  35 & 03:31:45.09 & 27:36:25.25 & 19.38 & 0.044 & 19.74 & 0.036 & 19.42 & 0.053 & 0.05 \\
  36 & 03:31:46.20 & 27:36:26.22 & 19.59 & 0.057 & 19.97 & 0.044 & 19.57 & 0.062 & 0.00 \\
  37 & 03:31:44.05 & 27:36:26.97 & 18.61 & 0.034 & 18.86 & 0.023 & 18.70 & 0.028 & 0.02 \\
  38 & 03:31:28.79 & 27:36:29.49 & 19.81 & 0.073 & 21.41 & 0.084 & 19.87 & 0.074 & 0.00 \\
  39 & 03:31:40.31 & 27:36:40.08 & 19.01 & 0.042 & 19.20 & 0.027 & 19.07 & 0.038 & 0.18 \\
  40 & 03:31:44.32 & 27:36:40.36 & 19.06 & 0.050 & 19.40 & 0.031 & 19.09 & 0.039 & 0.01 \\
\hline

\end{tabular}
\end{table*}

\newpage

\begin{figure}
\centerline{\hbox{
\psfig{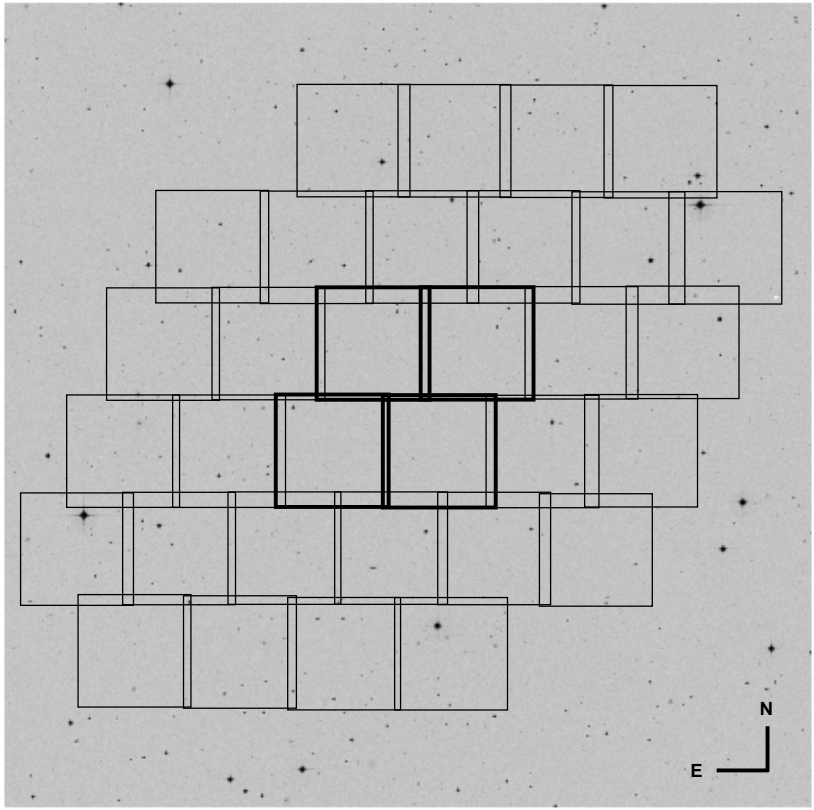}}}
\caption{The 32 SofI fields overlaid on a 35\arcmin square
Digitized Sky Survey image of the Chandra Deep Field South. 
Each SofI field is 4.9\arcmin\ square; the deeper central fields
are indicated by thicker outlines. \label{fig-pretty} }
\end{figure}

\newpage

\begin{figure}
\centerline{\hbox{
\psfig{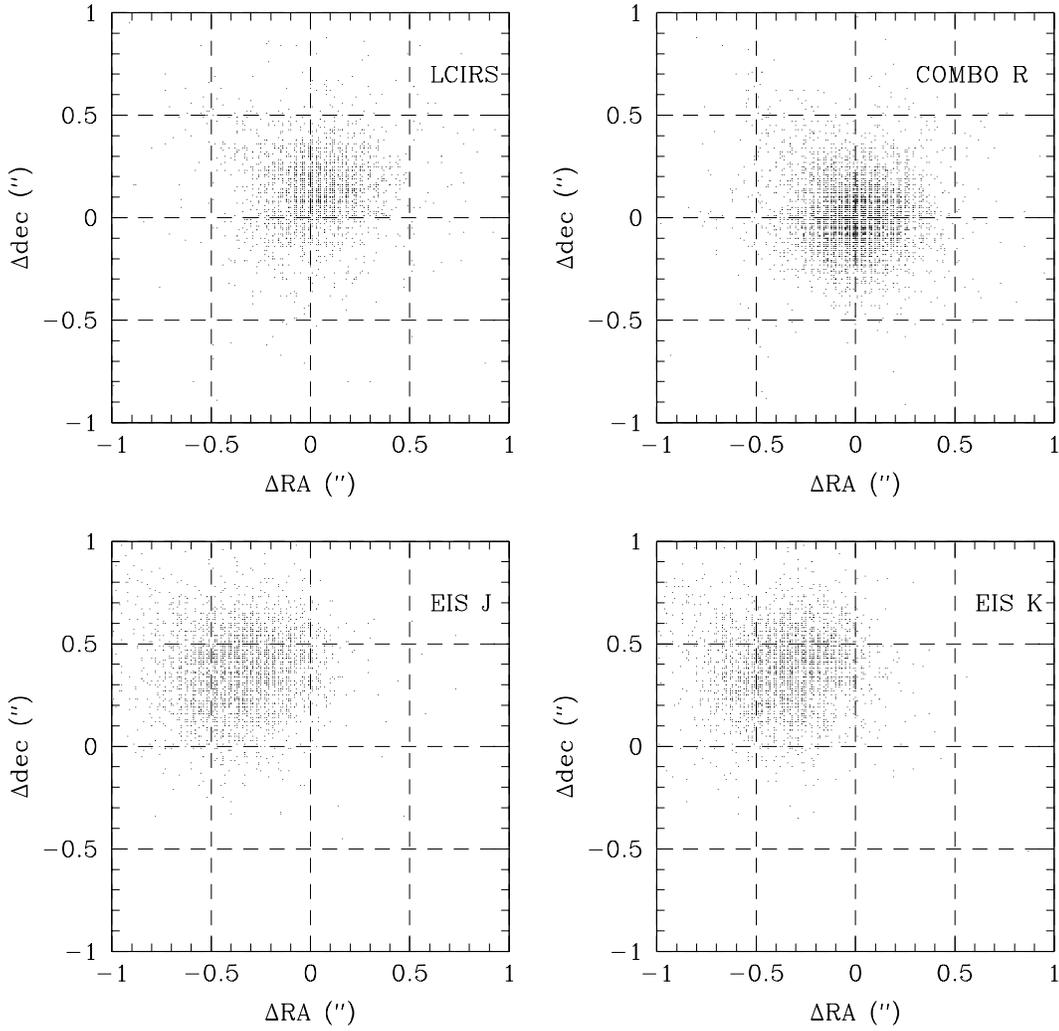}}}
\caption{Comparison between the coordinates of detections in common with the Las Campanas survey 
(top left), the $R$-band of the COMBO 17 survey (top right) and the $J$ and $K_s$-bands of the EIS survey 
(bottom). \label{astrometry}}
\end{figure}

\begin{figure}
\centerline{\hbox{
\psfig{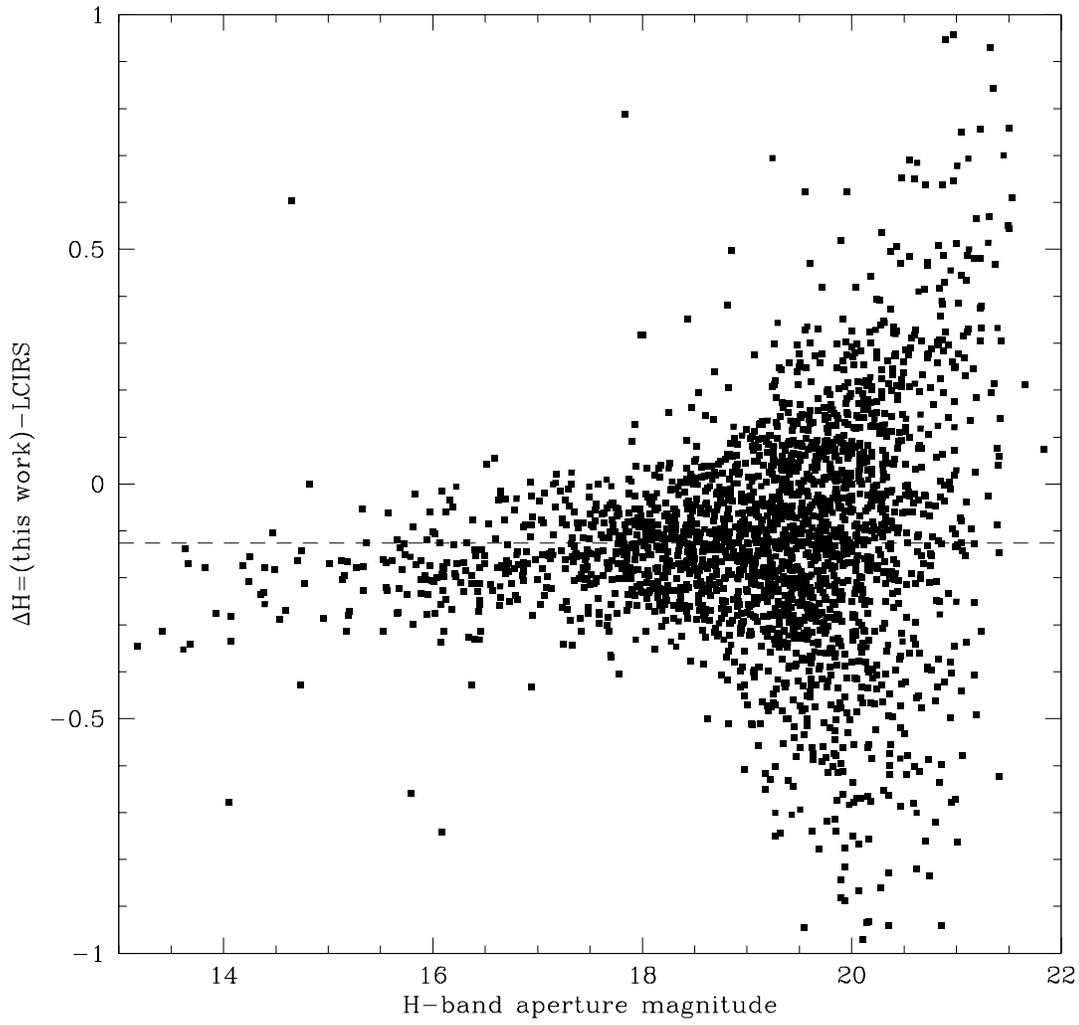}}}
\caption{Comparison of photometry between our data and that of LCIRS.
\label{photo}}
\end{figure}

\begin{figure}
\centerline{\hbox{
\psfig{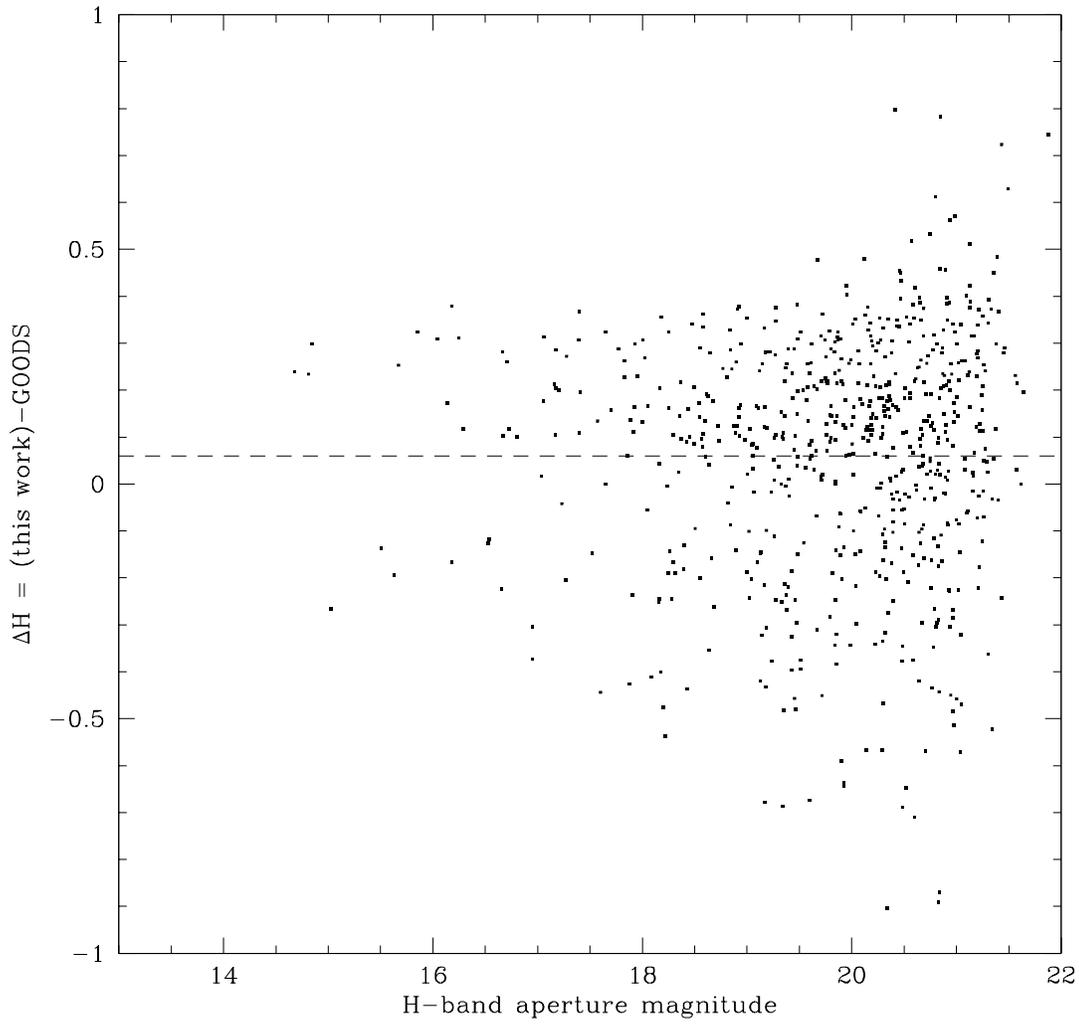}}}
\caption{Comparison of photometry between our data and that of 
GOODS \label{goods}}
\end{figure}

\begin{figure}
\centerline{\hbox{
\psfig{figure=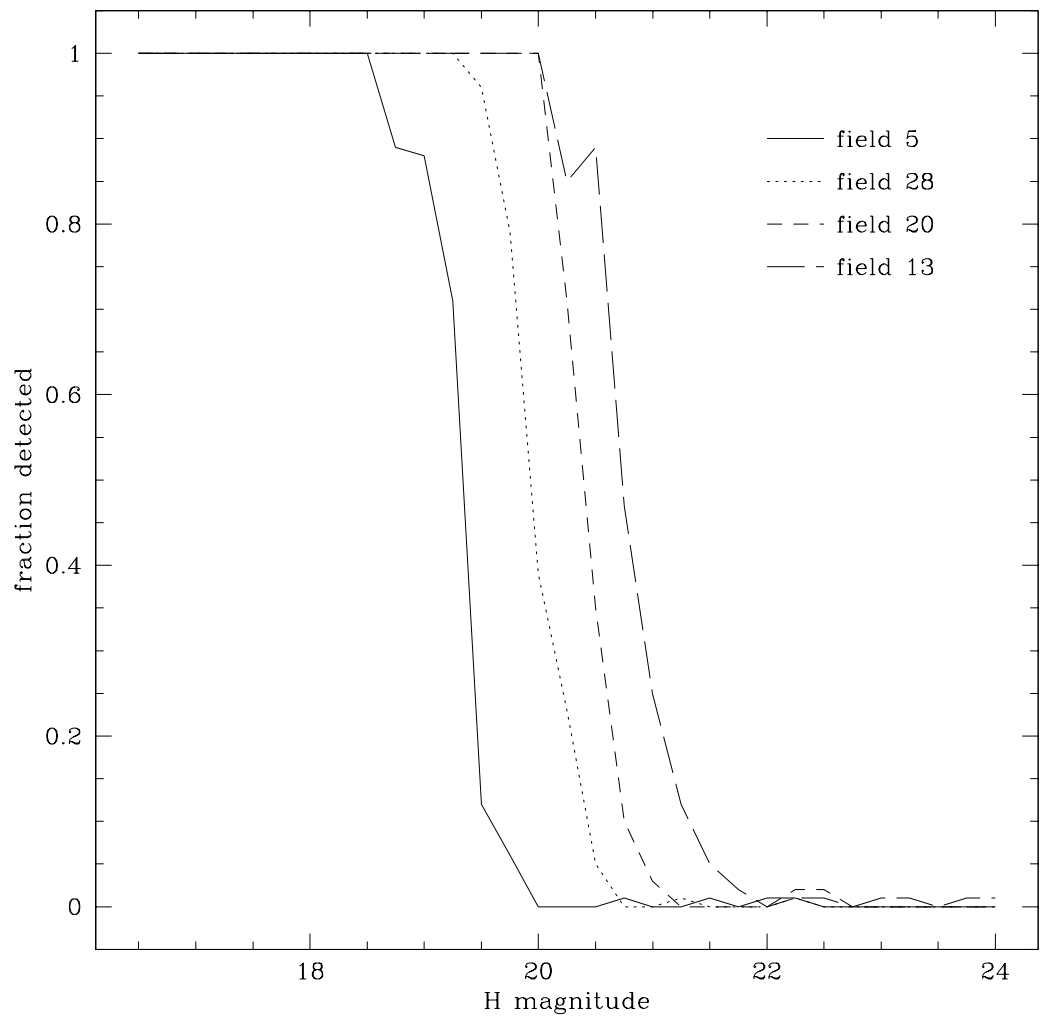,bbllx=25pt,bblly=170pt,bburx=565pt,bbury=690pt,width=14cm}}}
\caption{Completeness data for fields 5, 23, 13 and 19.\label{completeness}}
\end{figure}

\begin{figure}
\centerline{\hbox{
\psfig{figure=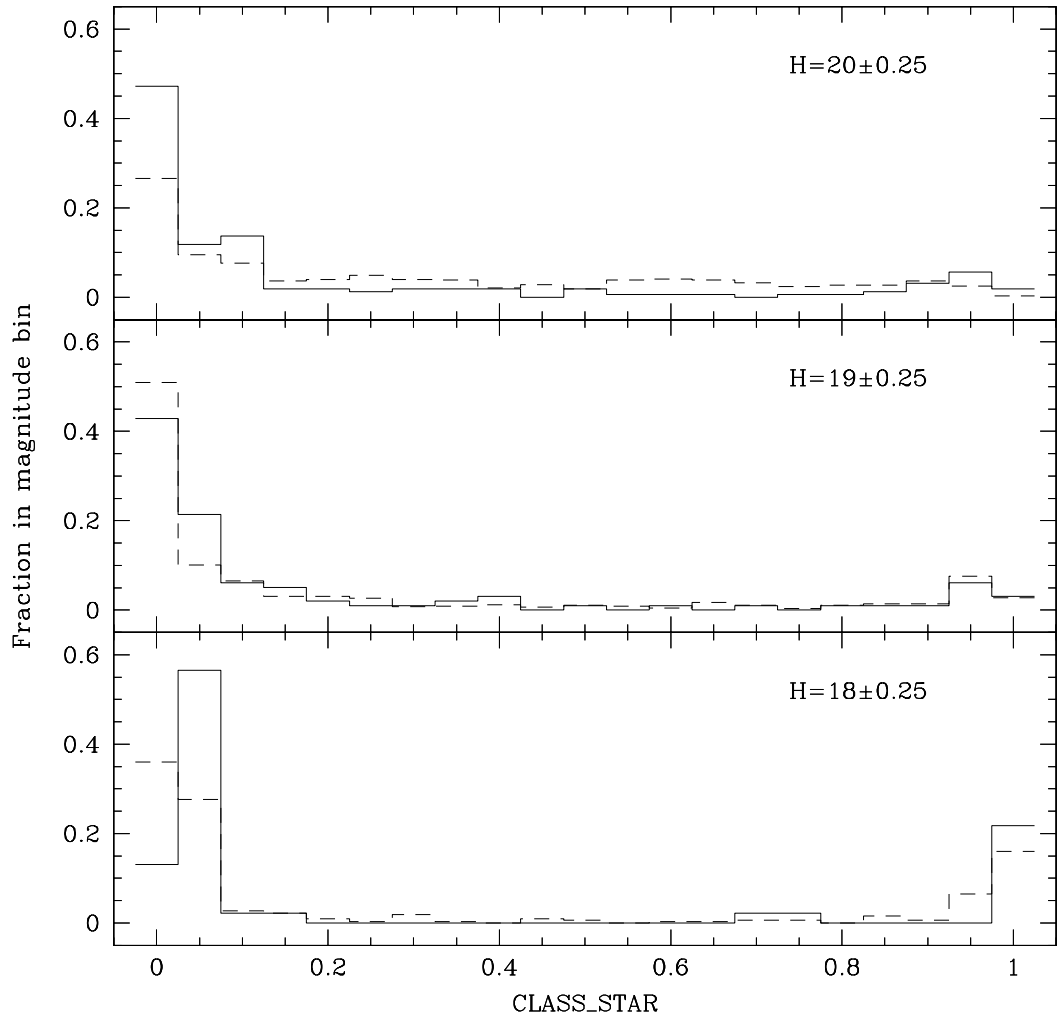,bbllx=25pt,bblly=170pt,bburx=565pt,bbury=690pt,width=14cm}}}
\caption{Distribution of CLASS\_STAR for objects in 3 different magnitude bins.\label{class_star}}
\end{figure}

\begin{figure}
\centerline{\hbox{
\psfig{figure=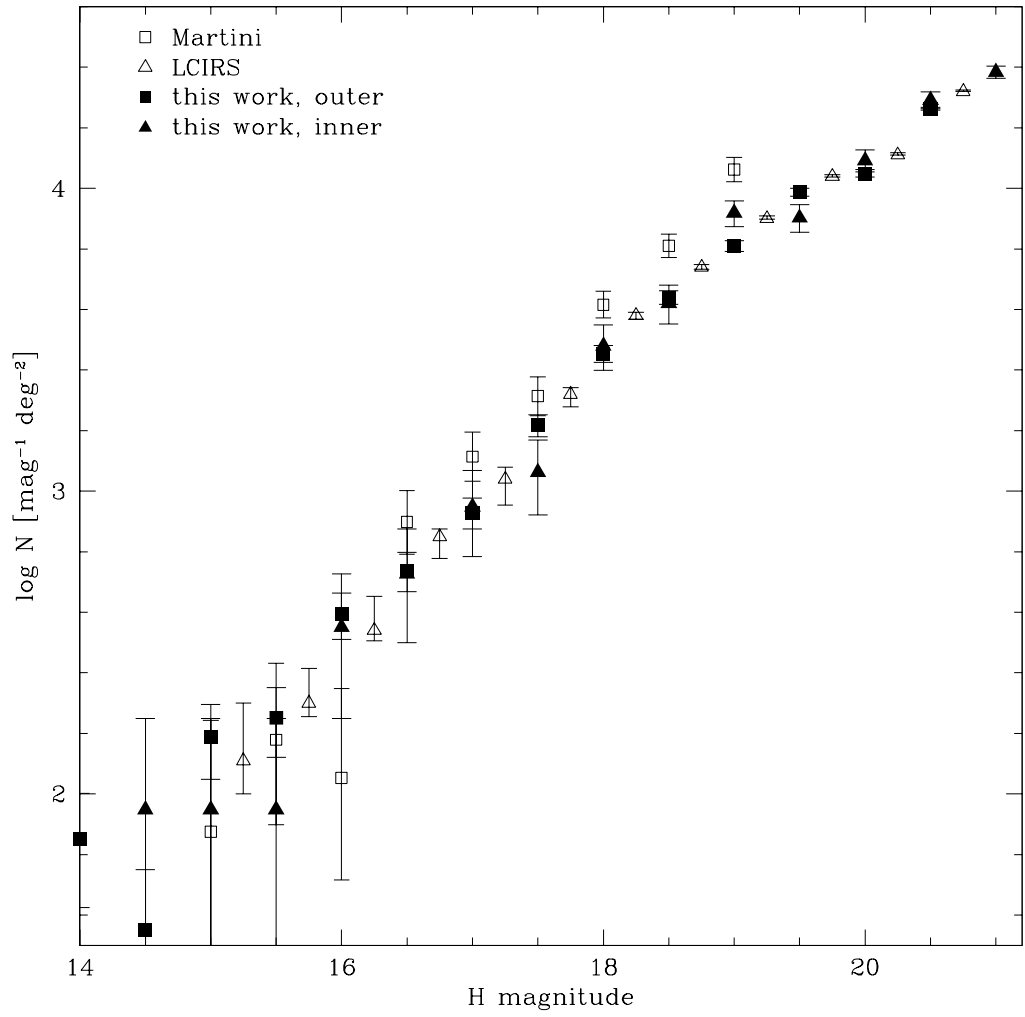,bbllx=25pt,bblly=170pt,bburx=565pt,bbury=690pt,width=14cm}}}
\caption{Galaxy number counts for the outer and inner fields. Open symbols
are data from \citet{lcirs3} and \citet{mar01}.\label{numc}}
\end{figure}

\end{document}